\documentclass[printer]{cDSS2e}

\usepackage{rotating,epsfig}
\begin{document}
\doi{10.1080/14689360xxxxxxxxxxxxx}
 \issn{1468-9375}
\issnp{1468-9367} \jvol{00} \jnum{00} \jyear{2006} \jmonth{June}

\markboth{Phase lock and  rotational motion of a parametric pendulum}{G. Litak, M. 
Borowiec,  
and M. Wiercigroch}

\title{
Phase lock and  rotational motion of a parametric pendulum in noisy and chaotic 
conditions}

\author{Grzegorz Litak$^1$, Marek Borowiec$^1$, and  Marian Wiercigroch$^2$ \\~\\
$^1$Department of Applied Mechanics, Technical University of
Lublin, \\
Nadbystrzycka 36, PL-20-618 Lublin, Poland \\
$^2$Centre for Applied Dynamics Research, Department of Engineering, \\ Fraser Noble 
Building,
       King's College, University of Aberdeen, \\ Aberdeen, AB24 3UE, Scotland, UK
}   \received{16 June 2006}

\maketitle

\begin{abstract}
The effect of noise on a rotational mode of a pendulum excited 
kinematically in vertical direction has been analyzed. 
We have shown that for a weak noise transitions from 
oscillations to rotations  
and vice versa are possible.
For a moderate noise level dynamics of the system is governed by a combination of the excitation 
amplitude and 
stochastic component. Consequently for stronger noise 
the rotational solution as an independent sychonized mode has vanished.
\end{abstract}

\section{Introduction}
A forced pendulum is one of the simplest nonlinear system which shows typical chaotic 
behaviour and this has been discussed extensively in many research articles and in several 
textbooks, for example see
\cite{Baker1996,Bardin1995,Moon1979,Szemplinska2000,Bishop1996,Szemplinska2002,Steindl1991,Leven1981,Smith1994,Litak1999,Xu2005}. 
It is worth noting the pendula have been studied not only to understand the fundamental nonlinear 
behaviour but also with a view to solve some practical problems of wave energy extraction from sea 
waves \cite{X2005}, gravitational gradient pendulum \cite{Suits2006,Ziegler2001}
and ship dynamics \cite{Falzarano1991}, to name a few. Some other nonlinear phenomena like the
Josephson junction \cite{Cicogna1987,Strogatz1994} and  can also be easily associated with the analysed
problem.

\begin{figure}[htb]
{\small (a)}
\epsfig{file=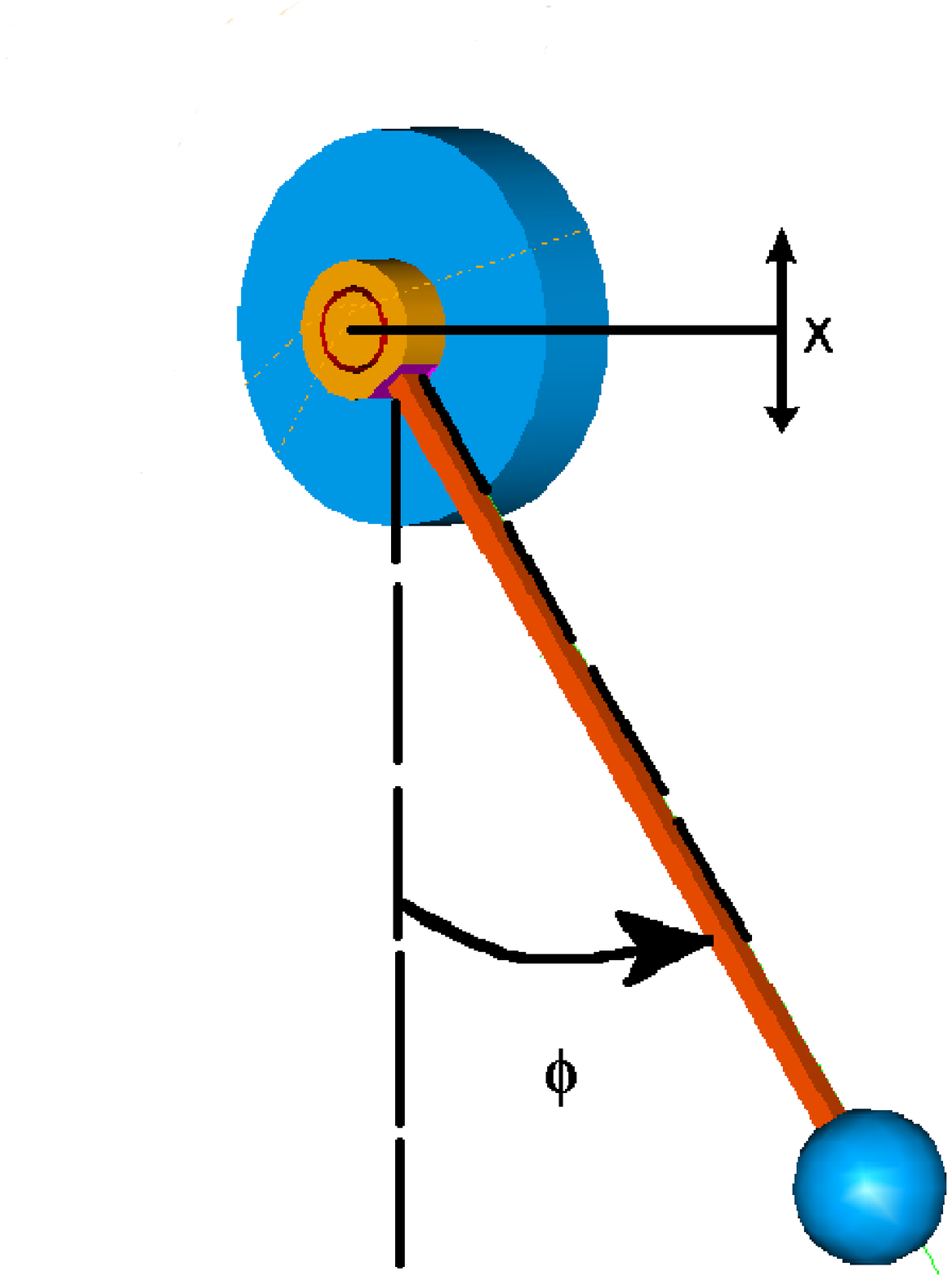,width=4.0cm,angle=0}
\vspace{-5cm}
~

\hspace{6cm}
\epsfig{file=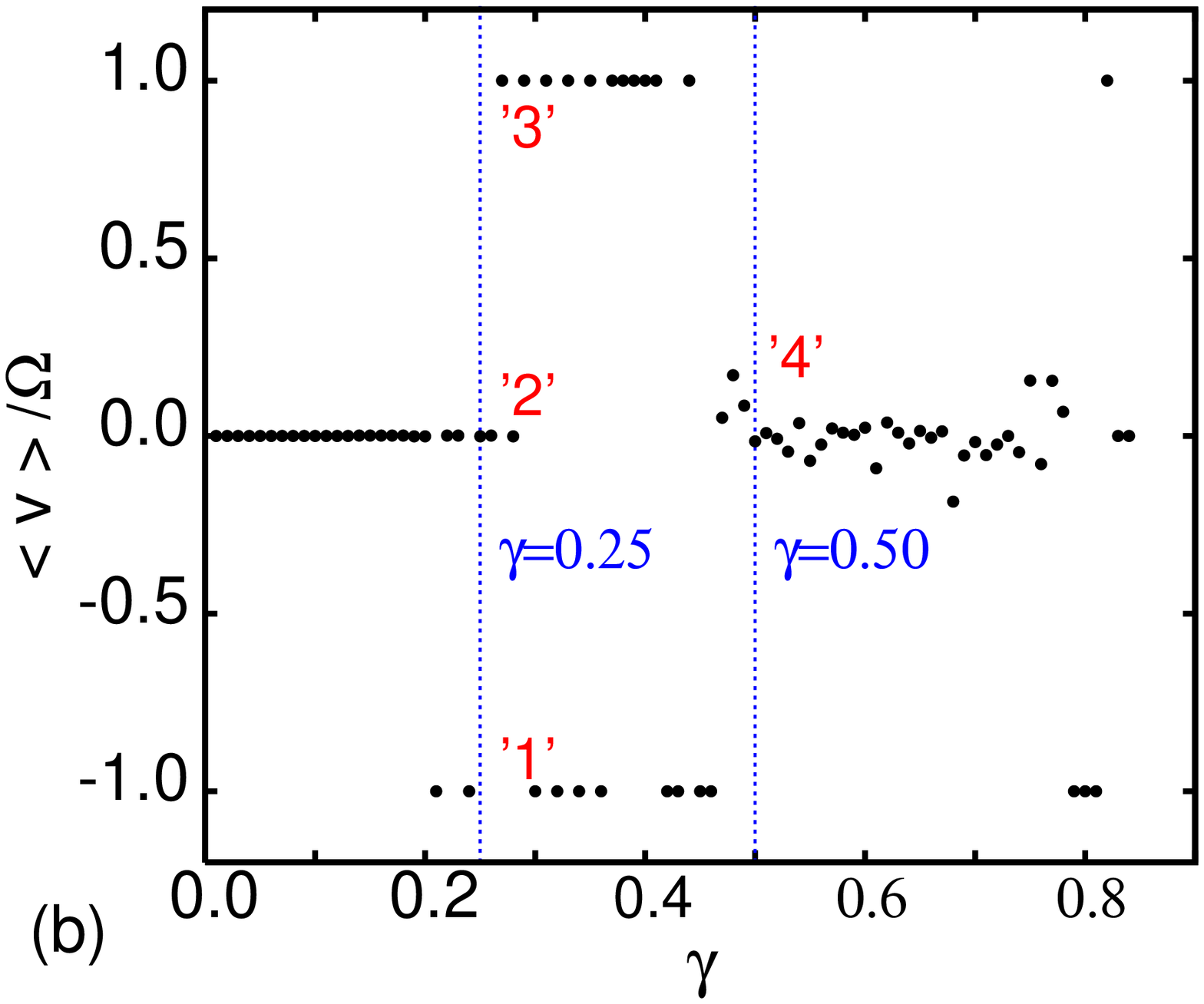,width=5.5cm,angle=-90}

\caption{ \label{fig1} (a) A parametric pendulum with harmonic vertical 
excitation of a 
suspension point 
$x=A \cos 
\Omega' t=A \cos
\Omega \tau$, (b) Rotational number $<v>/\Omega$ versus 
$\gamma$ for $\alpha=0.1$ ($v=\dot \phi$). Vertical lines correspond to $\gamma=0.25$ and 0.50
($\Omega=2$), 
respectively. 
Calculations 
have been 
done  starting  from the
smallest 
$\gamma \approx 0$ and increasing it up to $\gamma = 0.9$.
For each new $\gamma$ the initial conditions $\phi_0$ 
and $v_0$ were defined  by the final values of $\phi$ and $v$ for previous $\gamma$. Numbers:
'1', 
'2', '3' 
and 
'4' are related to synchronized motions: clockwise rotation, oscillation, anti-clockwise
rotation 
and
chaotic motion, respectively.}
\end{figure}

In this note we will examine stability of rotational 
motion of parametric
pendulum \cite{Szemplinska2002,Xu2005} in noisy conditions \cite{Blackburn2006}.
For a pendulum under the  harmonic
kinematic excitation (Fig. \ref{fig1}a) the equation of motion can be written as follows
\begin{equation}   
\label{eq1} 
\frac{{\rm d}^2 \phi}{{\rm d} t^2} + \frac{k}{ml^2} \frac{{\rm d} \phi}{{\rm d} t}
+ \left(\frac{g}{l}+ \frac{A}{l}\Omega'^2 \cos{\Omega' t}\right) \sin \phi
=0,
\end{equation}
where $m$ is point mass $l$ is a length of pendulum, $g$ is a gravitational constant. 
Introducing the natural frequency of free oscillations, $\omega_0=\sqrt{g/l}$ 
we define dimensionless time $\tau=\omega_0 t$ and  renormalized frequency
$\Omega$ $\Omega=\Omega'/\omega_0$.

\begin{figure}[htb]

\centerline{
\epsfig{file=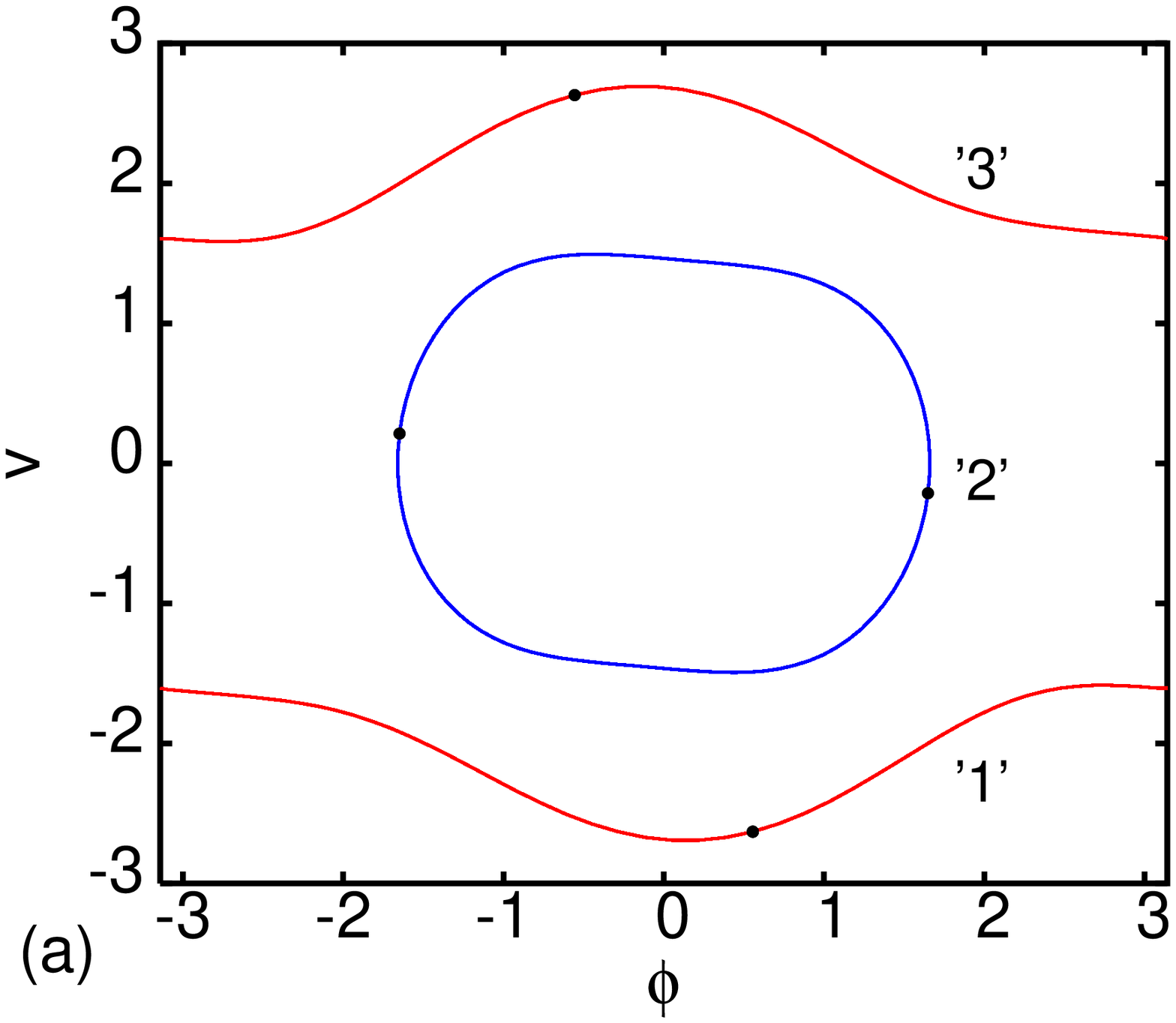,width=5.5cm,angle=-90}
\epsfig{file=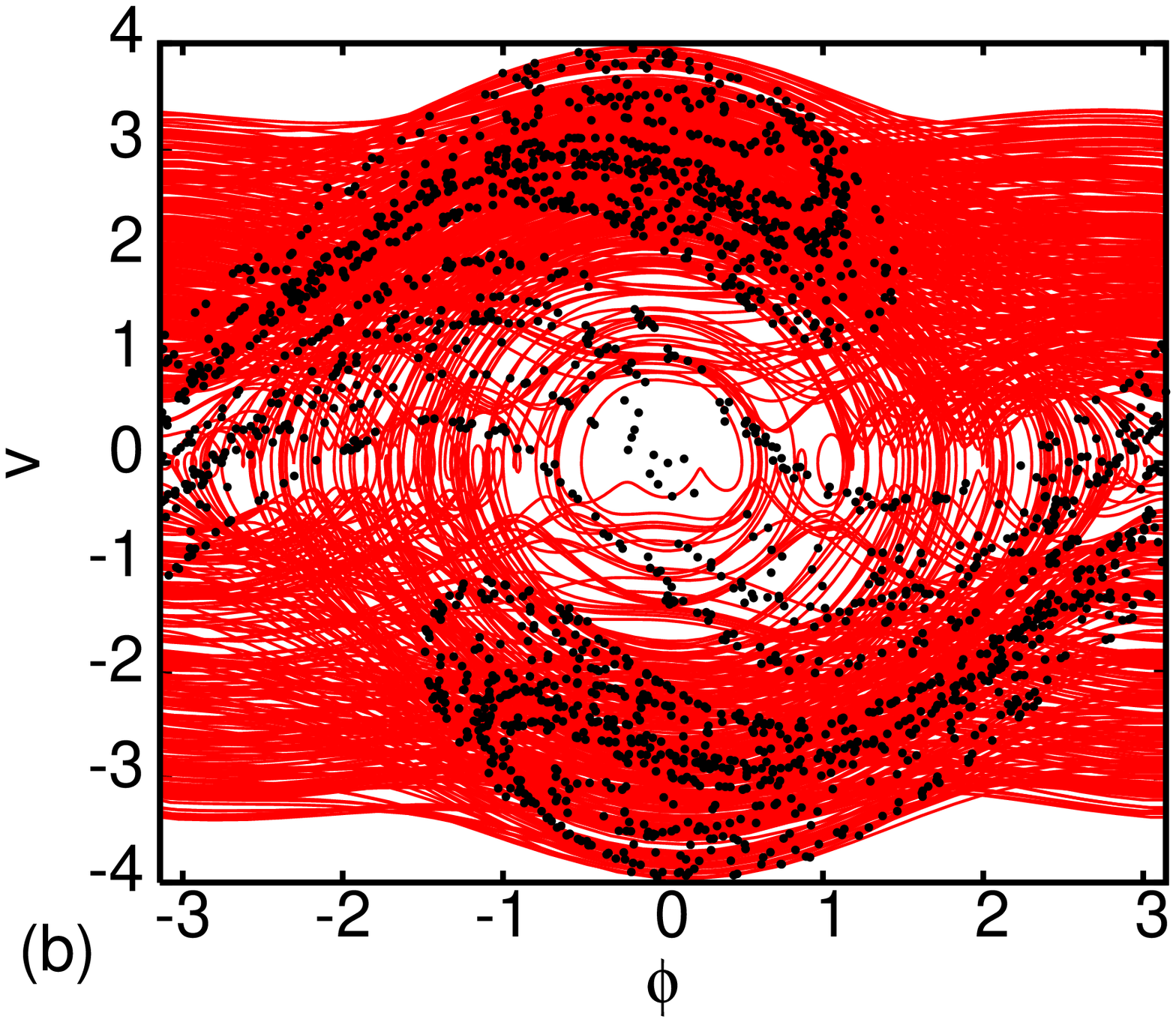,width=5.5cm,angle=-90}}

 \caption{ \label{fig2} Phase portraits and Poincare maps of the pendulum (Eq. 2) computed 
for $\omega=2$ $\gamma= 0.25$ and $\gamma=0.50$.  In Fig. \ref{fig2}a numbers '1', '2' and '3' correspond
to solutions marked 
in Fig. 
\ref{fig1}b ($\Omega=2$).}
\end{figure}

\begin{figure}[htb]
\centerline{
\epsfig{file=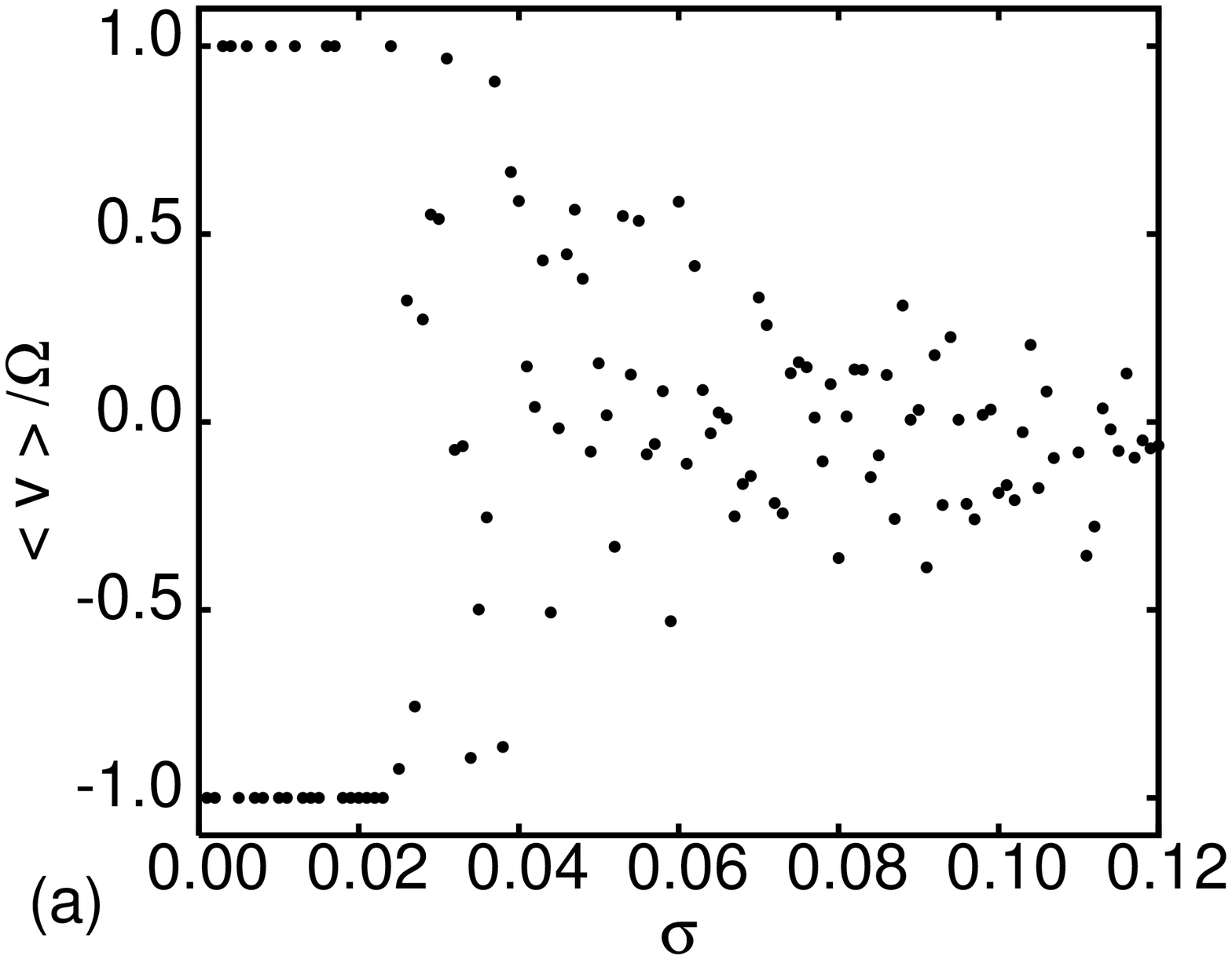,width=5.5cm,angle=-90}
\epsfig{file=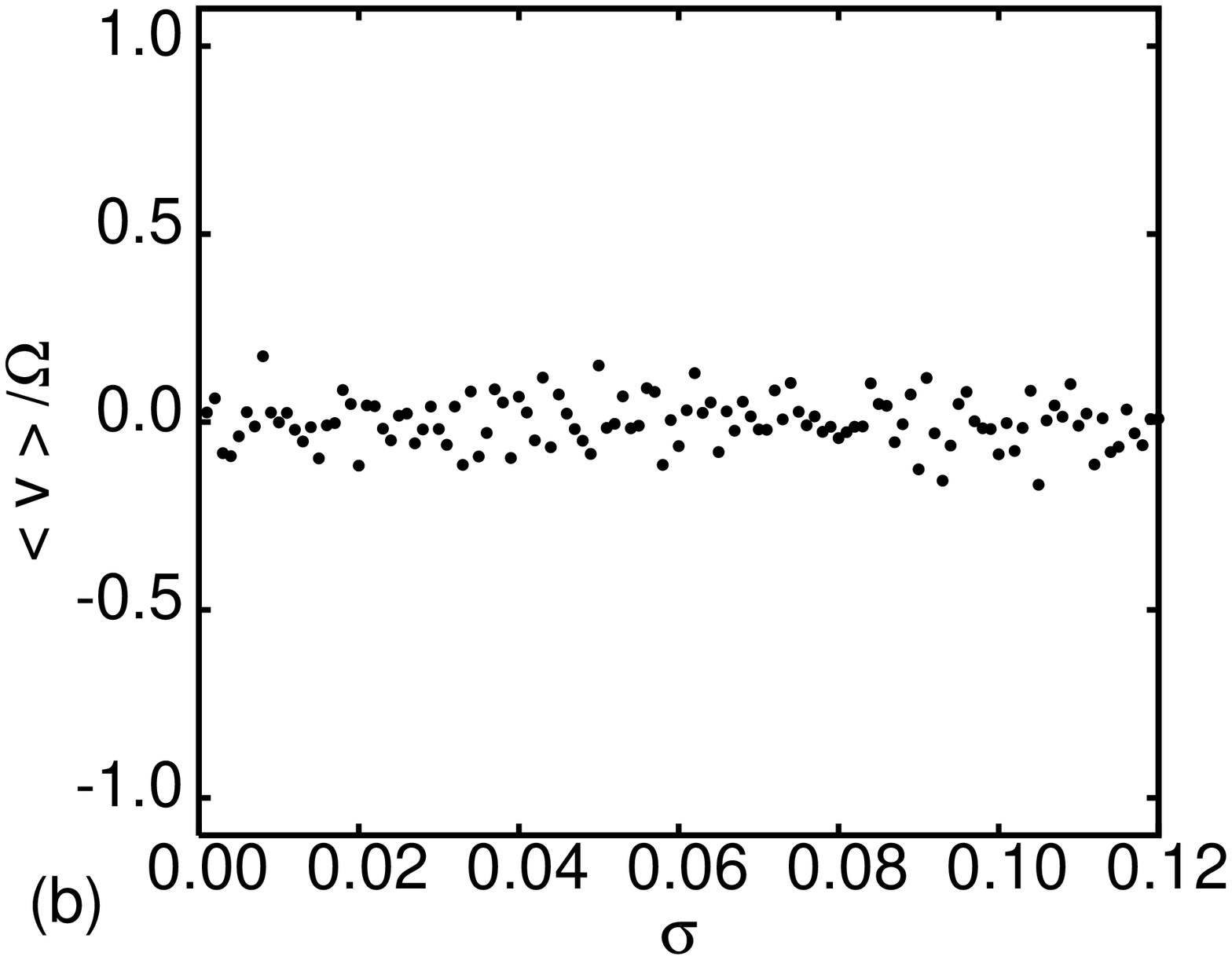,width=5.5cm,angle=-90}}
\centerline{
\epsfig{file=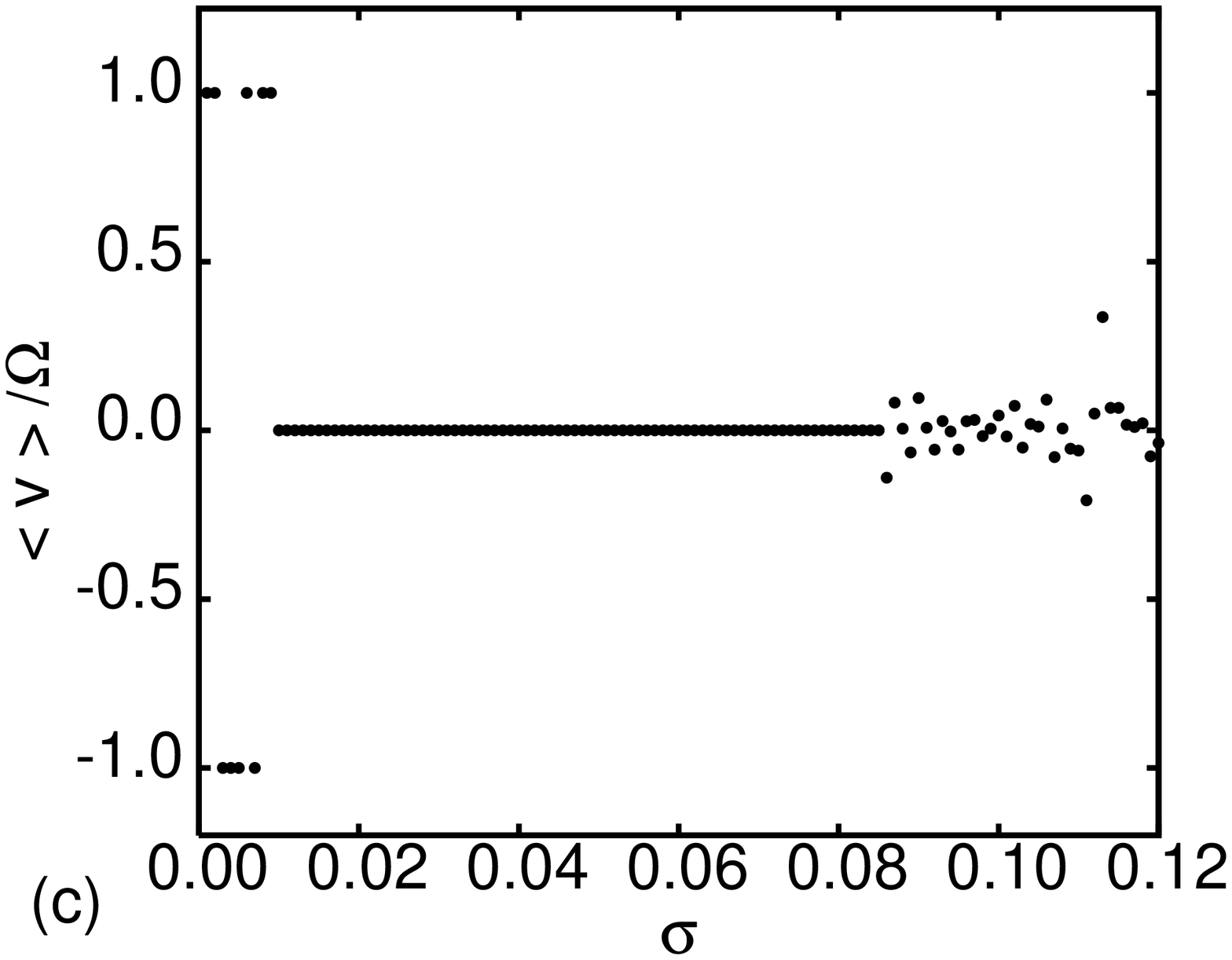,width=5.5cm,angle=-90}
\epsfig{file=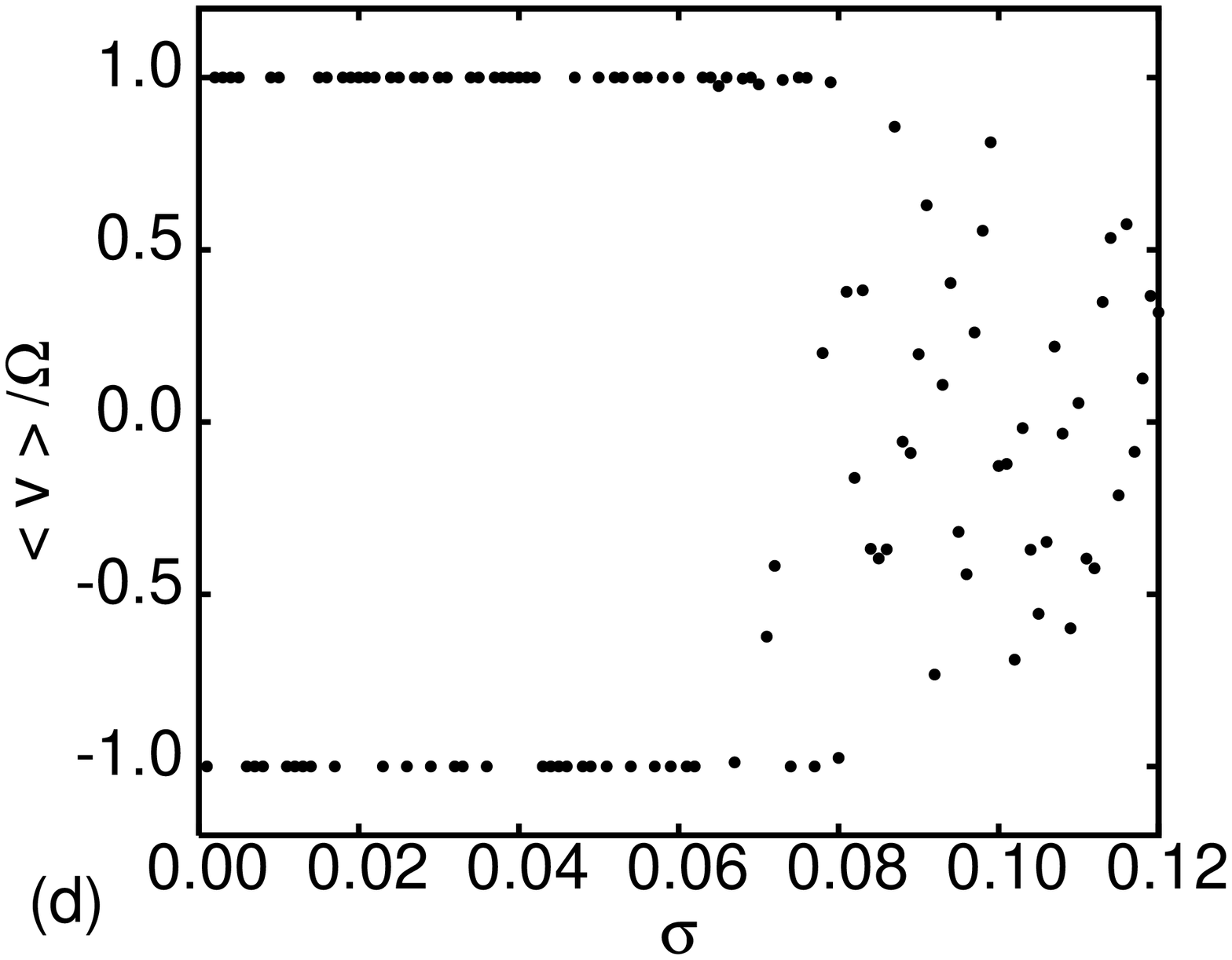,width=5.5cm,angle=-90}}

 \caption{ \label{fig3} Rotation number $< v> /\Omega$ ($\Omega=2$) versus noise level
$\sigma$ for 
$\alpha=0.1$ and four 
values of
$\gamma$; (a) $\gamma=0.4$, (b) $\gamma=0.6$,  
(c) $\gamma=0.8$,  
(d) $\gamma=1.0$. }
\end{figure}

\begin{figure}[htb]
\centerline{ 
\epsfig{file=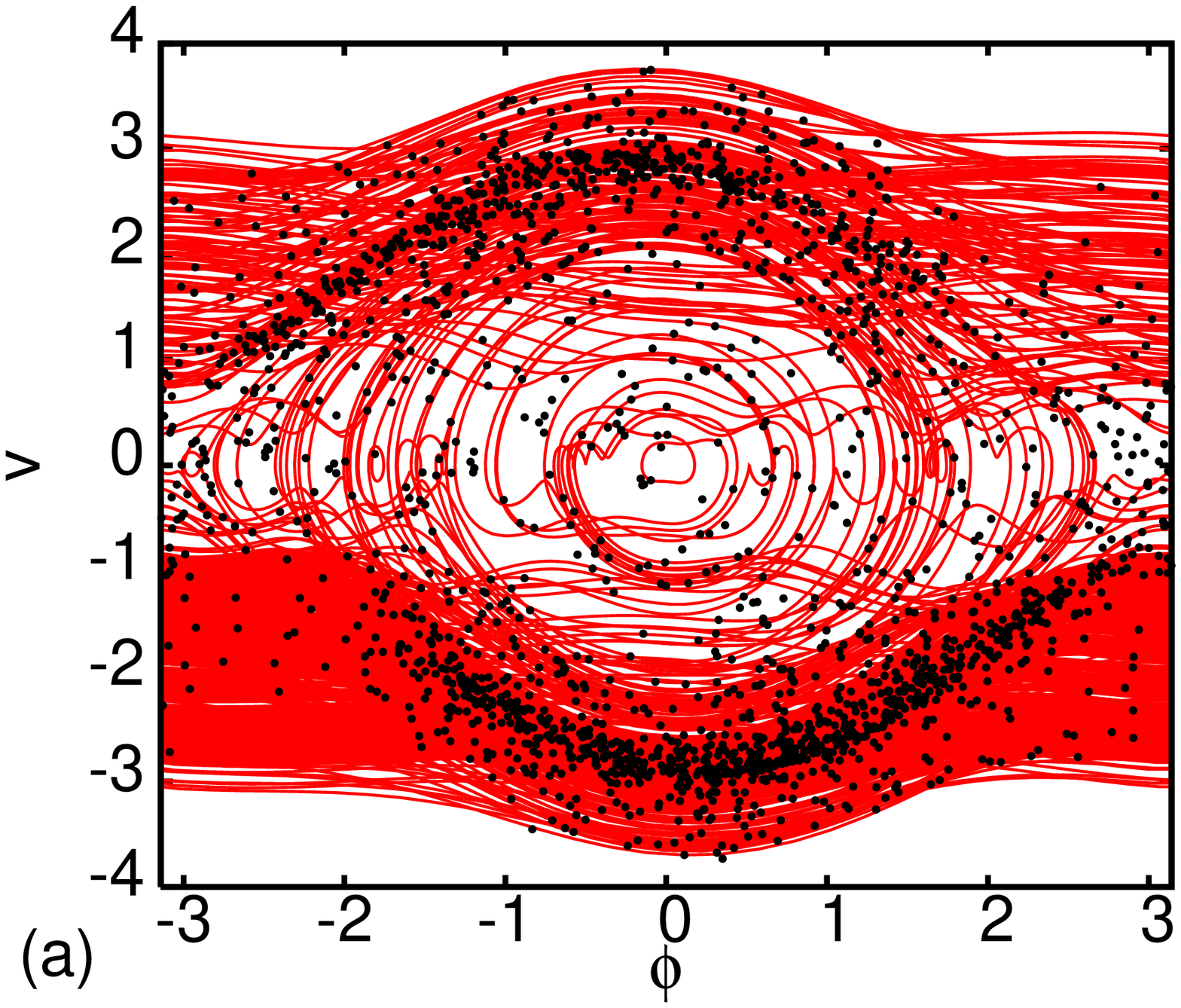,width=5.5cm,angle=-90}
\epsfig{file=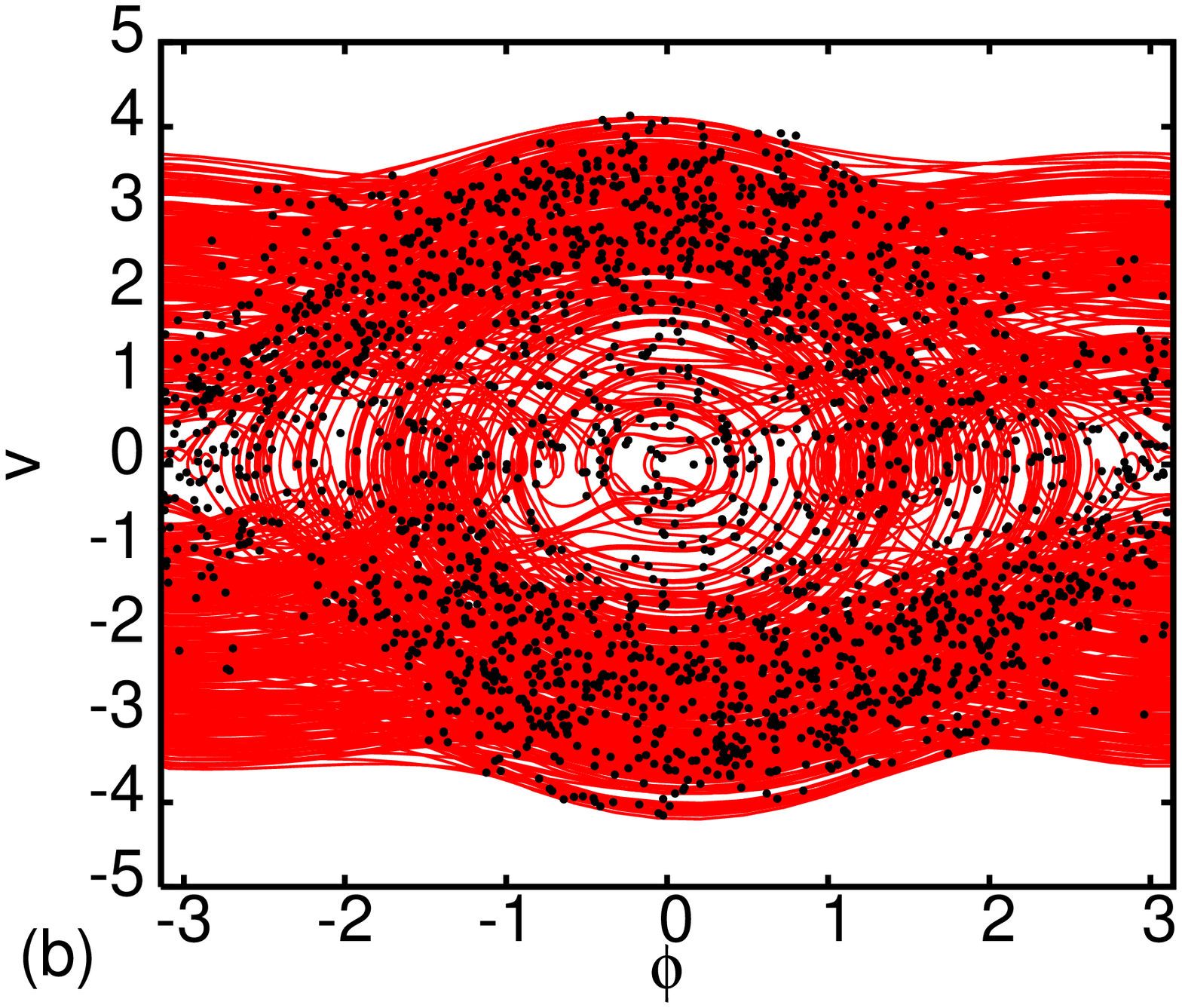,width=5.5cm,angle=-90}}
\centerline{
\epsfig{file=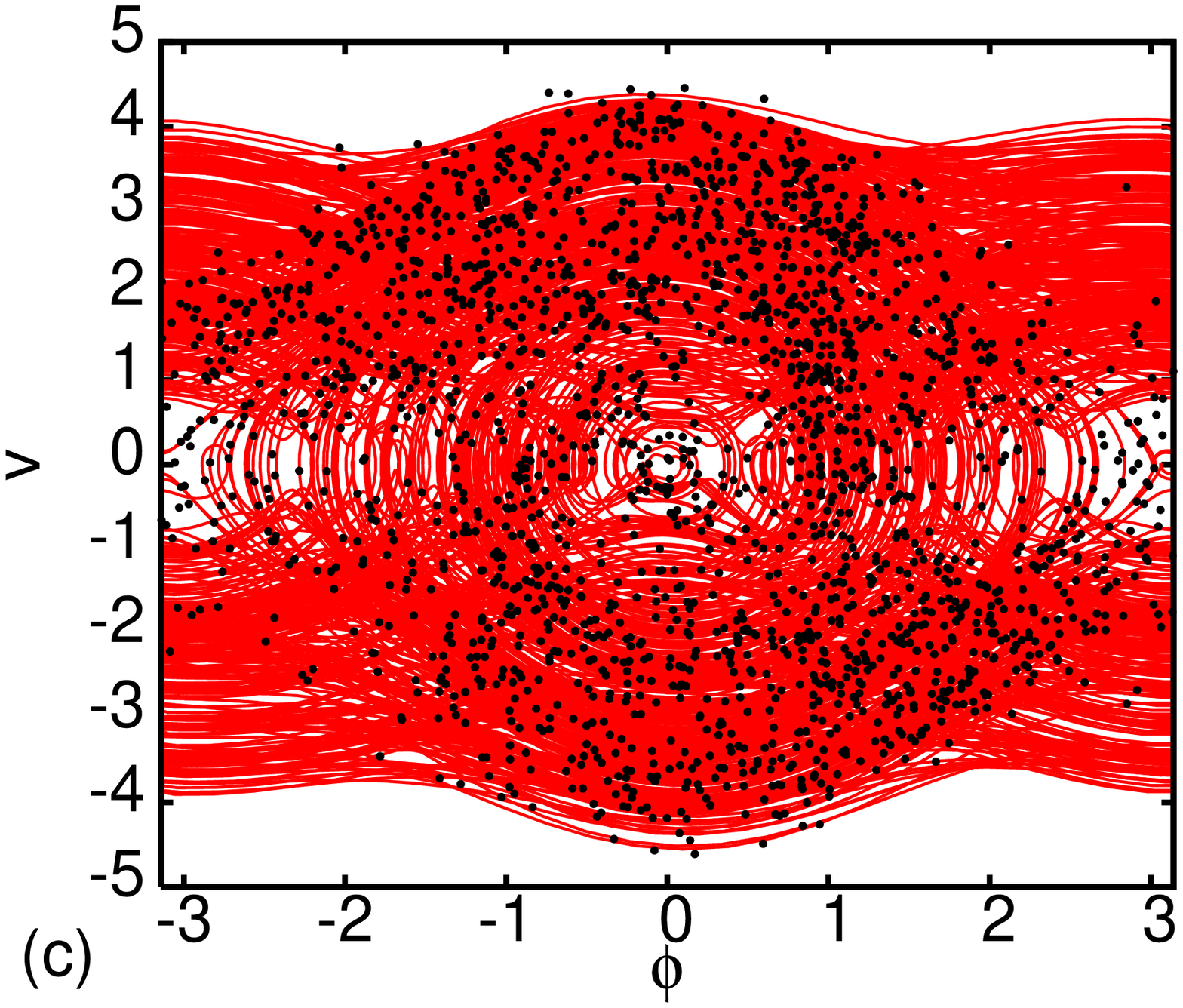,width=5.5cm,angle=-90}
\epsfig{file=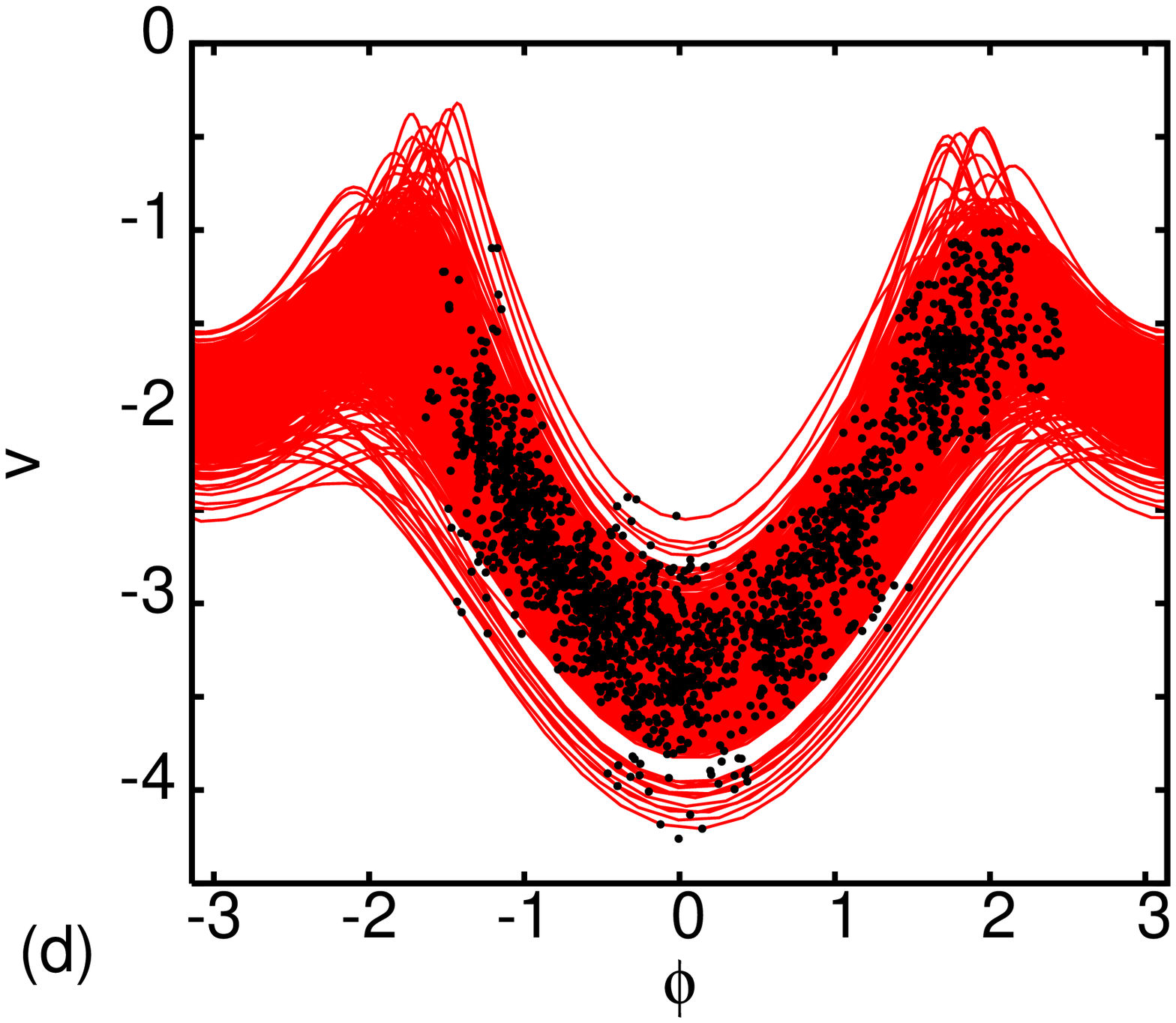,width=5.5cm,angle=-90}}

 \caption{ \label{fig4} Phase portraits and Poincare maps for   
$\alpha=0.1$, $\sigma=0.05$  and four
values of
$\gamma$; (a) $\gamma=0.4$, (b) $\gamma=0.6$, (c)  
$\gamma=0.8$,
(d) $\gamma=1.0$. Note different scales in $v$-axis.}

\end{figure}

\begin{figure}[htb]
\centerline{
\epsfig{file=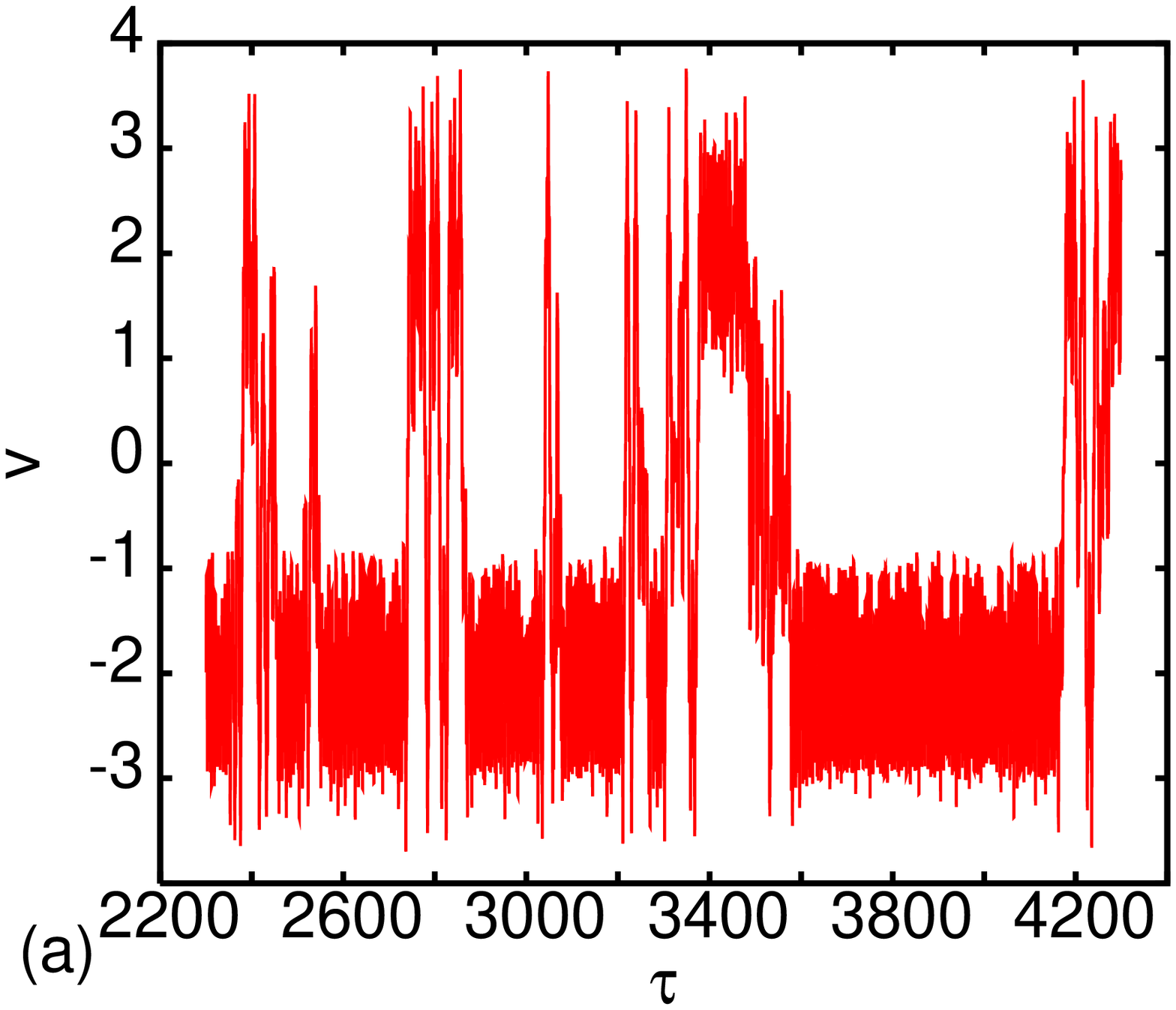,width=5.5cm,angle=-90}
\epsfig{file=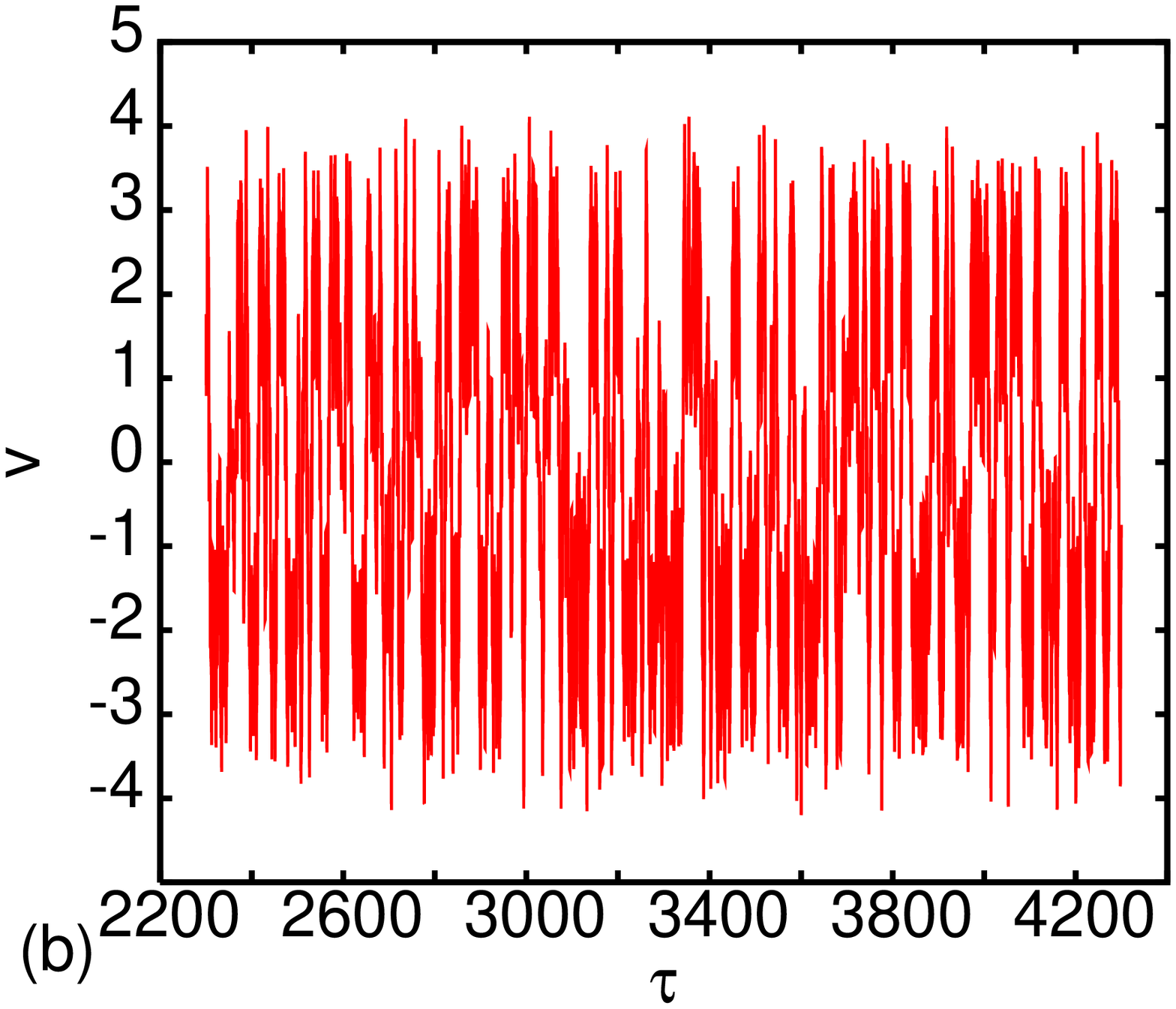,width=5.5cm,angle=-90}}
\centerline{
\epsfig{file=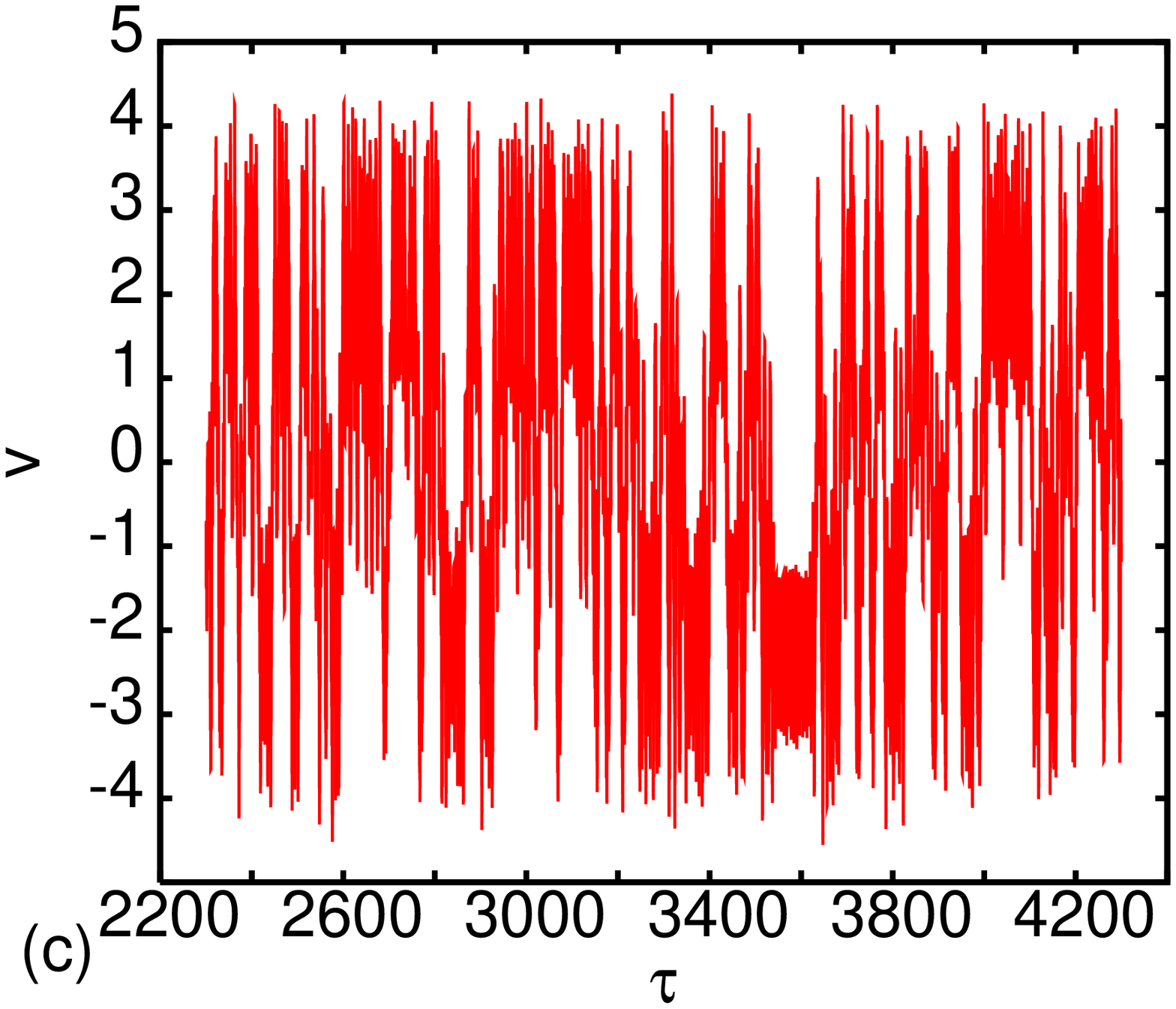,width=5.5cm,angle=-90}
\epsfig{file=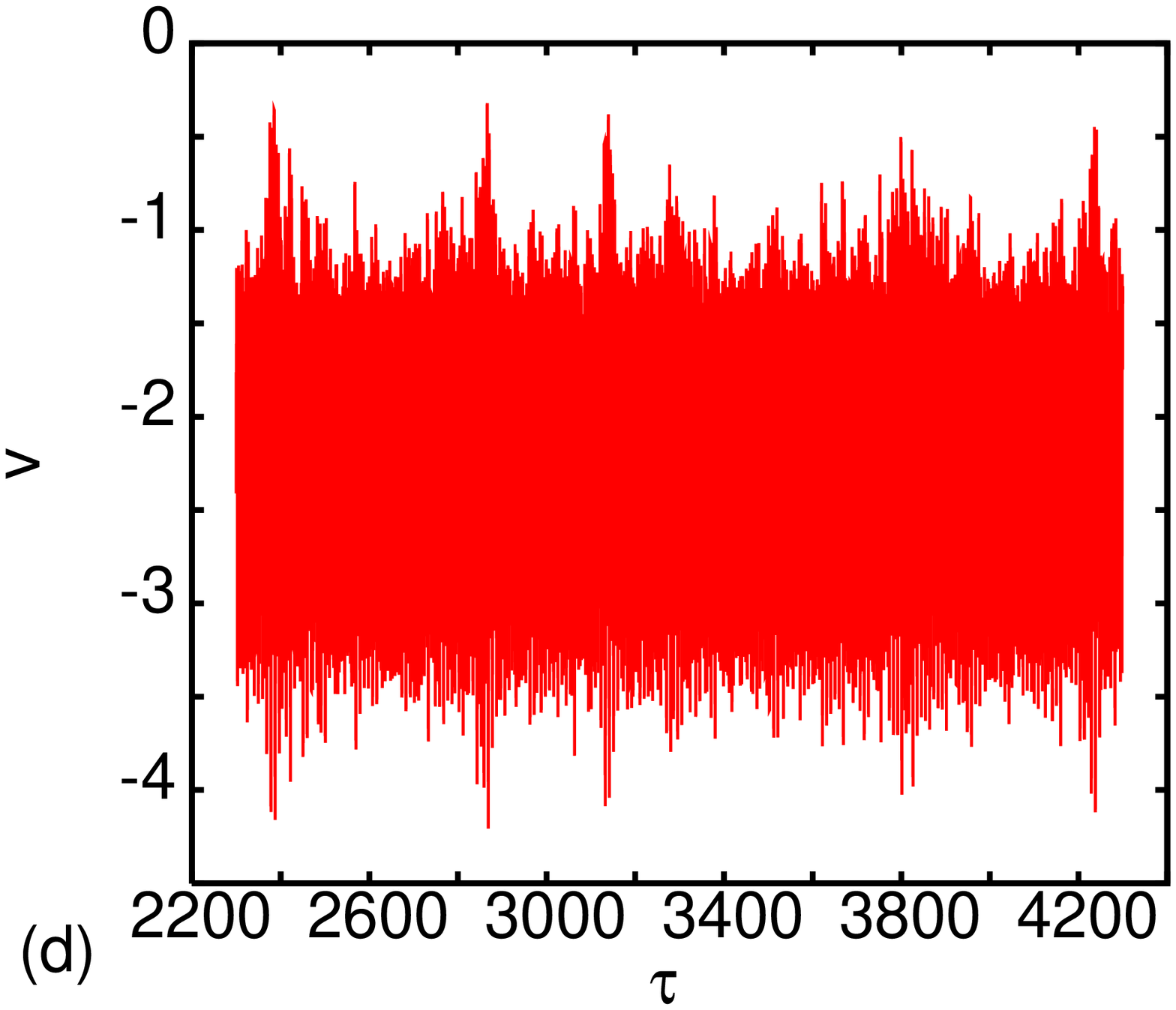,width=5.5cm,angle=-90}}

 \caption{ \label{fig5}  Time series showing the influence of the excitation strength, $\gamma$,
computed for
$\alpha=0.1$, $\sigma=0.05$, $\Omega=2$; (a)$\gamma=0.4$, 
(b)$\gamma=0.6$, (c)$\gamma=0.8$,
(d)$\gamma=1.0$. Note different scales in $v$-axis.}
\end{figure}

\begin{figure}[htb]
\centerline{
\epsfig{file=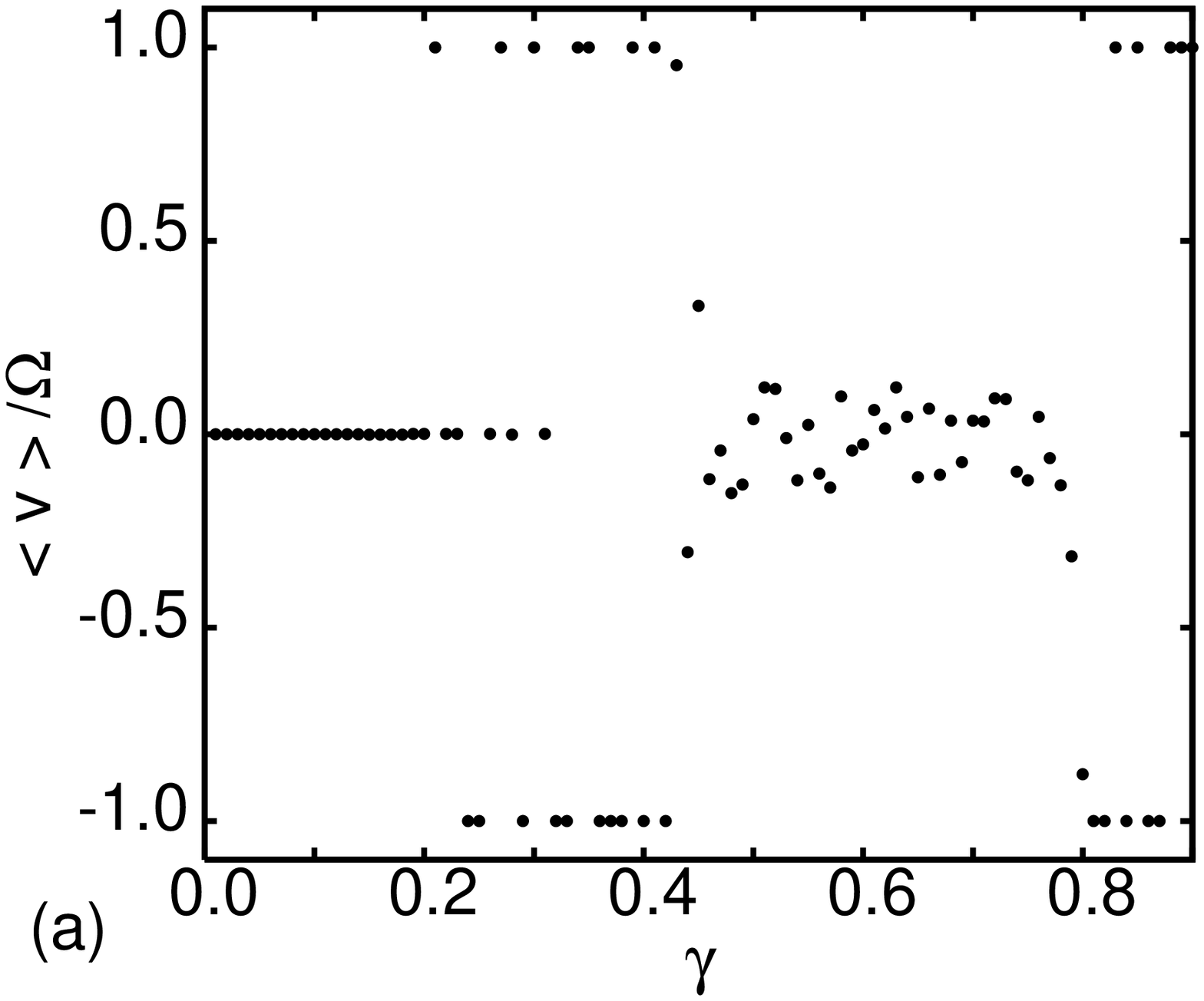,width=5.5cm,angle=-90}
\epsfig{file=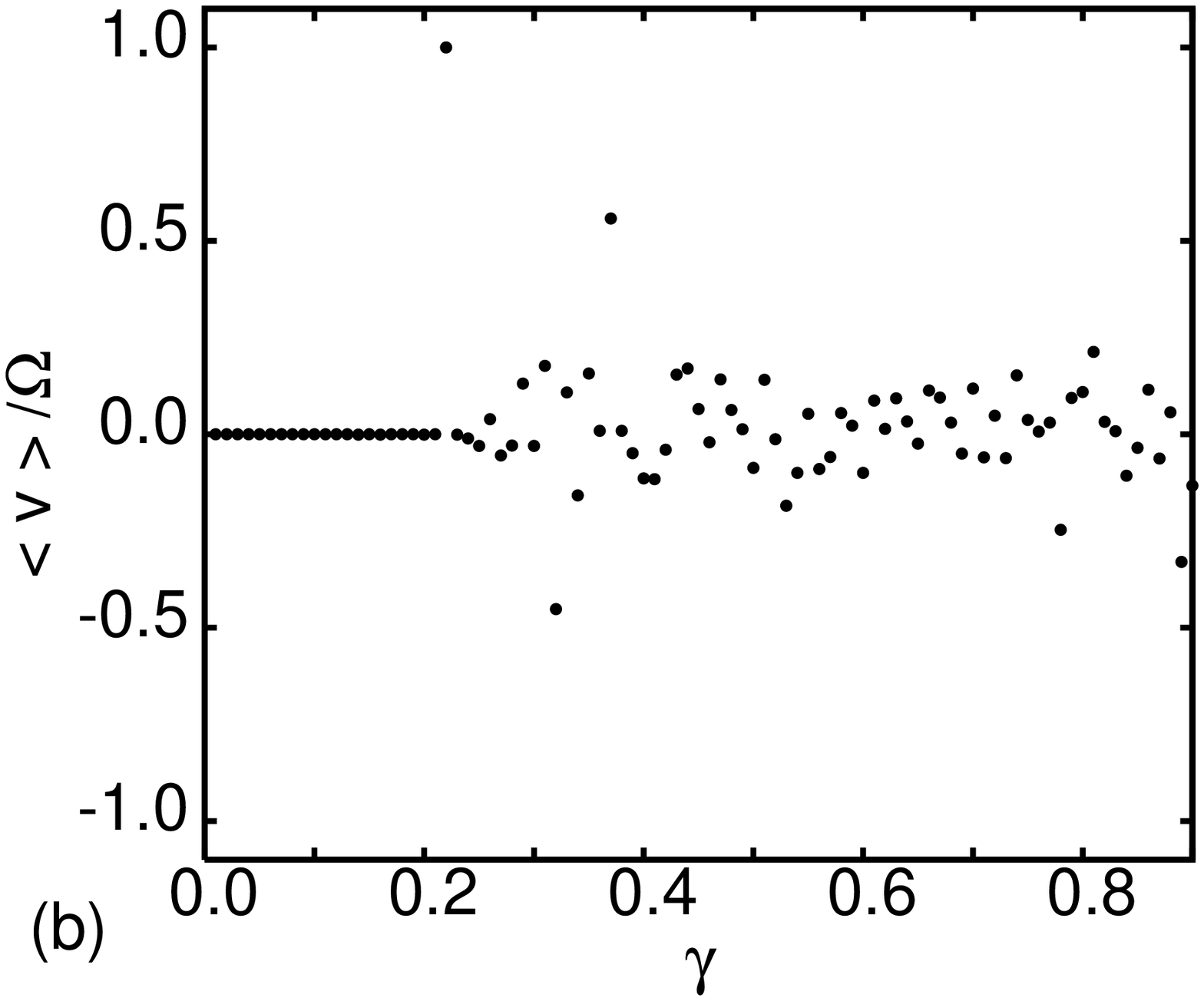,width=5.5cm,angle=-90}
}

 \caption{ \label{fig6} Rotational number $<v>/\Omega$ ($\Omega=2$) versus $\gamma$ for
$\alpha=0.1$ and two 
different noise levels; (a) $\sigma=0.01$, 
(b) $\sigma=0.1$.   
}
\end{figure}

Note the time dependent term $\frac{A\Omega^2}{l} \cos{\Omega \tau} \sin \phi$ is directly related
to
the inertial force due to the kinematic harmonic excitation with $\Omega'$  frequency (Fig.~\ref{fig1}a).
\begin{equation}
\label{eq2}
x(t)= A\cos\Omega' t =A\cos\Omega \tau.
\end{equation}

To simplify the notation and further analysis we also introduced two new parameters $\gamma=A/l$ 
and $\beta=k/(ml^2)$ while time derivative ${\rm d}/{\rm d} \tau \equiv \dot{}$  

\begin{equation}   
\label{eq3}
\ddot{\phi} + \beta \dot{\phi}
+ (1+ \gamma \Omega^2 \cos{\Omega \tau}) \sin \phi
=0.
\end{equation}

Deterministic system
regular oscillatory and rotational motions as well as chaotic oscillations can be easily classified
by observing changes in so called rotational number 
$<v>/\Omega$ defined in
\cite{Baker1996,Kim2004} as
\begin{equation}
\label{eq4}
<v>/\Omega = \lim_{\tau_2 \rightarrow \infty} \frac{1}{\Omega (\tau_2-\tau_1)} \int_{\tau_1}^{\tau_2} \dot
\phi ~{\rm d} \tau,
\end{equation}
where $\tau_1$ was the time interval necessary for the system to reach a steady state (here 
$\tau_1=2.3 \times 10^3$) while $\tau_2$ was chosen to be large enough  
($\tau_2=4.3 \times 10^3$). 
In Fig. \ref{fig1}b we show that this quantity versus $\gamma$ and one can clearly see the regions of
synchronized motions and their transitions into chaotic oscillations.  
Note for the synchronized motion the 
averaged rotational velocity is kept constant against $\gamma$ parameter up to $\gamma\approx 0.47$ where
we observe
different behaviour  (Fig. \ref{fig1}b). 
Namely  $<v>/\Omega =\pm 1$ 
for rotations 
and 0 
for
an oscillatory motion. The rational number of  $<v>/\Omega$ exhibits the phase lock
phenomenon
\cite{Baker1996} characteristic for nonlinear systems. The simulations of deterministic equation 
(Eq. \ref{eq5}) 
show that in 
the relatively large region of system 
parameters the average velocity is constant. Consequently pendulum motion can be decoupled into 
two modes: rotations 
with the average rotational velocity and superimposed oscillations.
To illustrate changes in corresponding attractors we plotted (in Fig. \ref{fig2}a and b) phase
portraits for 
$\gamma=0.25$ (and 
various initial conditions) and $\gamma=0.5$ representing different types of motions. 

In the next section we will examine the effect of noise in the examined system (Eq. \ref{eq3}).  In 
particular, we will discuss the simulation results of
the rotational  number in chaotic and noisy conditions, 
focusing on destabilization of rotational motions.
\newpage
\section{Phase lock in presence of noise}

To examine the system further 
we have transformed the second order differential equation into two equations of the first order
\begin{eqnarray}
\dot \phi &=& v
\\
\dot v  &=& -\beta \dot{\phi} \nonumber
- (1+ \gamma \Omega^2 \cos{\Omega \tau}) \sin \phi.
\label{eq5}
\end{eqnarray}
We assumed that in the system effected by noise through a kinematic excitation is not perfectly
periodic
resulting 
in variations of forcing frequency 
$\Omega'$ or its related period $2 \pi/\Omega'$ in time. 
Having this in mind we decided to mimic this effect by a bounded noise concept \cite{Liu2001} 
This type of noise is introduced by a random phase to kinematic forcing term $\cos{(\Omega
\tau+\Psi) }$,
there $\Psi=\Psi(\tau)$ is a Wiener process.
To perform  numerical calculations we have to discretise  our system 
(Eq. \ref{eq3}).
First we write  the  Euler-Maruyama  scheme of integration \cite{Naess2000}
\begin{eqnarray}   
v_{n+1} &=&  \left(-\beta v_n
- (1+ \gamma \Omega^2 \cos{\Theta_n}) \sin \phi_n \right) \Delta \tau \nonumber \\
\phi_{n+1} &=& v_n \Delta \tau  \label{eq6} \\
\Theta_{n+1} &=& \Theta_n + \Omega \Delta \tau + \Gamma(\tau,\sigma) \sqrt{ \Delta \tau},
\nonumber
\end{eqnarray}
where $\Gamma(\tau)$ is a random number with a normal distribution and  the standard deviation
$\sigma$ 
referred here as the level of noise.
To improve precision of our numerical calculations we have used the  Runge-Kutta-Maruyama algorithm 

\cite{Naess2000} 
which treats more carefully  the deterministic part of  Eq. \ref{eq6}.
Figures \ref{fig3}a--d show the rotational numbers $< v> /\Omega$ versus $\sigma$ for
different $\gamma$. 
Note that
in all cases rotational motion is destructed by noise, and large enough $\sigma$ the resulting cases
have variable 
average velocity. In Fig. \ref{fig3}b the deterministic state is originally chaotic for 
deterministic case and hence does 
not change in noisy conditions. Namely the result correspond to the chaotic oscillations with a noise
component. 
For better 
clarity we have also plotted the phase portraits and Poincare maps for one the chosen value of
$\sigma=0.5$
and the same values of $\gamma$ ($\gamma=0.4$, 0.6, 0.8, and  1.0 for Fig. \ref{fig4}a--d 
respectively). Note, first 
three
figures (Fig. \ref{fig4}a--c) show the system with large 
amplitude variations but not synchronized rotations;
this appears to be chaotic oscillations without any clear synchronized rotation
$< v> /\Omega\approx 0$.  On the other hand Fig. \ref{fig4}d is representing 
clock-wise rotations with the fixed average $< v> /\Omega=-1$.      
The corresponding time series are presented in Figs. \ref{fig5}a--d. 
Phase lock phenomenon is present in Fig. \ref{fig5}d where we observe anti-clockwise 
rotation (persistent negative velocity sign).
Surprisingly the succeeding Poincare points are smeared by random excitations leaving average
velocity 
$<v>/\Omega=-1$ 
unchanged.   
In Fig. \ref{fig5}a, one can see intermittent rotation with a negative sign.  
Note that in this case for $\sigma=0$ the system behaves regularly (Fig. \ref{fig3}a)  
with clear 
clockwise or 
anti-clockwise rotations ( $<v>/\Omega =\pm 1$). Consequently one can identify a path
to chaotic oscillations induced by
noise.
 Finally, in Figs. \ref{fig6}a and b we show  $<v>/\Omega$ versus $\gamma$ for two 
different levels of noise;
$\sigma=0.01$ for  Fig. \ref{fig6}a and $\sigma=0.1$ for Fig. \ref{fig6}b, respectively.
For $\sigma=0.01$ (Fig. \ref{fig6}a) we observe a similar sequence of transitions from oscillatory 
to rotational 
motions 
as in the deterministic case (Fig. \ref{fig1}b),  while  the noise level, $\sigma=0.1$, as shown in
Fig. \ref{fig6}b, is strong enough to destroy the synchronized  rotations for any $\gamma$.

\section{Summary and Conclusions}

We have performed numerical analysis of a parametrically excited pendulum
focussing on the rotational motion and
the influence of noise in the chaotic regimes. The rotational motion has
been computed for the pure deterministic
system and for the case with weak bounded noise introduced through a random
phase in the excitation
term. We found out, that in a large interval of increasing noise level the
average rotational velocity was stable; the noise component
created oscillations around the rotational average velocity. However, for a
large enough noise level $\sigma = 0.1$ (Fig. \ref{fig6}a) the rotational
motion as an independent sychonized mode has vanished.

Looking for a rotational motion of pendulum we observed an intermittent
transition to chaotic motion induced by noise. Such a conclusion was also
given
by Blackburn in his recent paper \cite{Blackburn2006} where the effect of potential
well escape was analyzed assuming a random external force. In our case
motivated by a possibility of
sea wave energy extraction by a parametric pendulum device, we considered
stochastic distribution
of the wave phase in succeeding periods.

\section*{Acknowledgements}
This research has been partially supported by the 
6th Framework Programme, Marie Curie Actions, Transfer of Knowledge, Grant No. 
MTKD-CT-2004-014058.

\end{document}